\documentclass[twocolumn,showpacs,preprintnumbers,amsmath,amssymb,prb]{revtex4}

\usepackage{graphicx}
\usepackage{dcolumn}
\usepackage{bm}
\usepackage{tabularx}
\usepackage{amsmath}

\begin{document}

\title{Thermal conductivity of overdoped BaFe$_{1.73}$Co$_{0.27}$As$_2$ single crystal: Evidence for nodeless multiple superconducting gaps and interband interactions}

\author{J. K. Dong,$^1$ S. Y. Zhou,$^1$ T. Y. Guan,$^1$ X. Qiu,$^1$ C. Zhang,$^1$ P. Cheng,$^2$ L. Fang,$^2$ H. H. Wen,$^2$ S. Y. Li$^{1,*}$}

\affiliation{$^1$Department of Physics, Surface Physics Laboratory (National Key Laboratory), and Laboratory of Advanced Materials, Fudan University, Shanghai 200433, China\\
$^2$National Laboratory for Superconductivity, Institute of Physics
and Beijing National Laboratory for Condensed Matter Physics,
Chinese Academy of Sciences, Beijing 100190, China}

\date{\today}

\begin{abstract}
The in-plane thermal conductivity $\kappa$ of overdoped FeAs-based
superconductor BaFe$_{1.73}$Co$_{0.27}$As$_2$ ($T_c$ = 8.1 K) single
crystal was measured down to 80 mK. In zero field, the residual
linear term $\kappa_0/T$ is negligible, suggesting a nodeless
superconducting gap in the $ab$-plane. In magnetic field,
$\kappa_0/T$ increases rapidly, very different from that of
conventional $s$-wave superconductors. This anomalous
$\kappa_0/T(H)$ may reveal an exotic superconducting gap structure
in overdoped BaFe$_{1.73}$Co$_{0.27}$As$_2$: the vanishing hole
($\beta$) pocket has a much larger gap than the electron ($\gamma$
and $\delta$) pockets which contain most of the carriers. Such an
exotic gap structure is an evidence for superconducting state
induced by interband interactions, in which the band with the {\it
smaller} density of states has a {\it larger} gap.
\end{abstract}


\maketitle

\section{Introduction}

For the recently discovered FeAs-based high-$T_c$ superconductors,
\cite{Kamihara,XHChen,GFChen,ZARen,RHLiu} the pairing symmetry of
its superconducting gap remains the most important issue to resolve.
Although extensive experimental and theoretical work have been done,
there is still no consensus. \cite{KIshida,Mazin1} While
angle-resolved photoemission spectroscopy (ARPES) experments clearly
demonstrated nearly isotropic multi-gaps,
\cite{HDing,TKondo,LZhao,LWray,KNakayama,KTerashima} the Andreev
spectroscopy, \cite{TYChien,LShan,Gonnelli} NMR,
\cite{KMatano,HGrafe,KAhilan,MYashima} and penetration depth
experiments
\cite{CMartin1,KHashimoto,LMalone,CMartin2,RGordon,TJWilliams} gave
conflicting claims on whether there are nodes in the superconducting
gaps.

Low-temperature thermal conductivity measurement is a powerful bulk
tool to probe the superconducting gap structure. \cite{Shakeripour}
The residual linear term $\kappa_0/T$ is very sensitive to the
existence of gap nodes and the field dependence of $\kappa_0/T$ can
give useful information on multi-gaps. Very recently, several heat
transport studies have been done on this new family of FeAs-based
and related superconductors. For the hole-doped
Ba$_{1-x}$K$_x$Fe$_2$As$_2$ ($T_c \simeq$ 30 K) \cite{XGLuo} and
electron-doped BaFe$_{1.9}$Ni$_{0.1}$As$_2$ ($T_c$ = 20.3 K),
\cite{LDing} a negligible $\kappa_0/T$ was found in zero field,
indicating a full superconducting gap. By contrast, a large
$\kappa_0/T$ was observed in BaFe$_{1.86}$Co$_{0.14}$As$_2$.
\cite{YMachida} For the prototype FeSe$_x$ ($T_c$ = 8.8 K)
superconductor, the thermal conductivity shows clear behavior of
multiple nodeless superconducting gaps. \cite{JKDong} In two
superconductors with lower $T_c$, $\kappa_0/T$ of BaNi$_2$As$_2$
($T_c$ = 0.7 K) is consistent with a dirty fully gapped
superconductivity, \cite{NKurita} while LaFePO ($T_c$ = 7.4 K)
appears to have a finite $\kappa_0/T$, suggesting the gap on some
band may have nodes. \cite{MYamashita}

For the most interested hole- and electron-doped BaFe$_2$As$_2$
superconductors, all samples studied by heat transport so far are
near optimal doping. \cite{XGLuo,LDing,YMachida} It will be very
interesting to study highly underdoped and overdoped samples to
demonstrate its superconducting gap structure over the whole doping
range. Furthermore, due to the high $T_c$ and $H_{c_2}$ of optimally
doped sampls, magnetic field can only be applied up to about 30\% of
their $H_{c_2}$. While for highly underdoped and overdoped samples
with relatively lower $T_c$, one may get a complete $\kappa_0/T(H)$
behavior to see if it has the multigap character, as in FeSe$_x$.
\cite{JKDong}

In this paper, we measure the thermal conductivity $\kappa$ of a
highly overdoped BaFe$_{1.73}$Co$_{0.27}$As$_2$ single crystal with
$T_c = 8.1$ K down to 80 mK. In zero field, the residual linear term
$\kappa_0/T$ is negligible, suggesting a nodeless superconducting
gap, at least in $ab$-plane. In magnetic field, $\kappa_0/T(H)$
increases sharply, very different from the
Ba$_{1-x}$K$_x$Fe$_2$As$_2$ and BaFe$_{1.9}$Ni$_{0.1}$As$_2$ samples
near optimal doping. Such an anomalous $\kappa_0/T(H)$ likely
results from an exotic superconducting gap structure in overdoped
BaFe$_{1.73}$Co$_{0.27}$As$_2$: the vanishing hole ($\beta$) pocket
has a much larger gap than the electron ($\gamma$ and $\delta$)
pockets which contain most of the carriers. Our finding of this
exotic gap structure supports the $s_{\pm}$-wave superconducting
state induced by interband interactions in FeAs-based
superconductors.

\section{Experimental}

Single crystals with nominal formula BaFe$_{1.7}$Co$_{0.3}$As$_2$
were prepared by self flux method. \cite{LFang} The diamagnetic
superconducting transition was measured by a vibrating sample
magnetometer (VSM) based on a physical property measurement system
(PPMS of Quantum Design) with the magnetic field perpendicular to
the $ab$-plane of the crystals. Energy Dispersive of X-ray (EDX)
microanalysis (Hitach S-4800) show that the actual Co content is
0.27. The sample was cleaved to a rectangular shape of dimensions
2.1 $\times$ 1.4 mm$^2$ in the $ab$-plane, with 25 $\mu$m thickness
along the $c$-axis. Contacts were made directly on the fresh sample
surfaces with silver paint, which were used for both resistivity and
thermal conductivity measurements. The contacts are metallic with
typical resistance 150 m$\Omega$ at 1.5 K. In-plane thermal
conductivity was measured in a dilution refrigerator down to 80 mK,
using a standard four-wire steady-state method with two RuO$_2$ chip
thermometers, calibrated {\it in situ} against a reference RuO$_2$
thermometer. Magnetic fields were applied along the $c$-axis and
perpendicular to the heat current. To ensure a homogeneous field
distribution in the sample, all fields were applied at temperature
above $T_c$.

\section{Results and Discussion}

\begin{figure}
\includegraphics[clip,width=7cm]{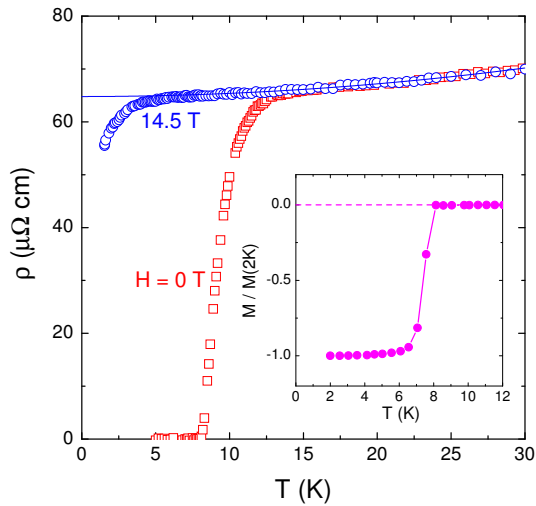}
\caption{(Color online) In-plane resistivity $\rho(T)$ of
BaFe$_{1.73}$Co$_{0.27}$As$_2$ single crystal in $H$ = 0 and 14.5 T.
The zero-resistance point of the resistive transition is at $T_c$ =
8.1 K in zero field. The solid line is a fit of the $H$ = 14.5 T
data between 10 and 30 K to the Fermi liquid form $\rho = \rho_0 +
AT^2$, which gives residual resistivity $\rho_0$ = 64.8 $\mu \Omega$
cm. Inset: normalized magnetization which shows the diamagnetic
superconducting transition.}
\end{figure}

\begin{figure}
\includegraphics[clip,width=7.15cm]{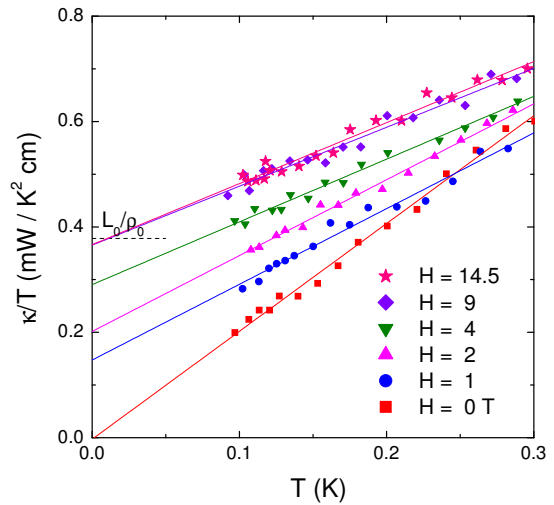}
\caption{(Color online) Low-temperature thermal conductivity of
BaFe$_{1.73}$Co$_{0.27}$As$_2$ in magnetic fields applied along the
$c$-axis ($H$ = 0, 1, 2, 4, 9, and 14.5 T). The solid lines are
$\kappa/T = a + bT$ fits (see text). The dashed line is the normal
state Wiedemann-Franz law expectation $L_0$/$\rho_0$, with $L_0$ the
Lorenz number 2.45 $\times$ 10$^{-8}$ W $\Omega$ K$^{-2}$.}
\end{figure}

Fig. 1 shows the in-plane resistivity of our
BaFe$_{1.73}$Co$_{0.27}$As$_2$ single crystal in $H$ = 0 and 14.5 T.
The zero-resistance point of the resistive transition is at $T_c$ =
8.1 K in zero field, which is consistent with the diamagnetic
superconducting transition shown in the inset of Fig. 1. The
residual resistivity $\rho_0$ = 64.8 $\mu \Omega$ cm is extrapolated
from the $H$ = 14.5 T data between 10 and 30 K, by using the Fermi
liquid form $\rho = \rho_0 + AT^2$.

To estimate the upper critical field $H_{c2}(0)$ which completely
suppresses the resistive transition, we define $T_c(onset)$ at the
temperature where $\rho(T)$ deviates from the $T^2$ dependence, and
get $T_c(onset)$ = 13.0 and 5.0 K for $H$ = 0 and 14.5 T,
respectively. Using the relationship $H_{c2}/H_{c2}(0) = 1 -
(T/T_c(0))^2$, $H_{c2}(0)$ = 17.0 T is obtained.

In Fig. 2, the temperature dependence of the in-plane thermal
conductivity for BaFe$_{1.73}$Co$_{0.27}$As$_2$ in $H$ = 0, 1, 2, 4,
9, and 14.5 T magnetic fields are plotted as $\kappa / T$ vs $T$.
All the curves are roughly linear, therefore we fit the data to
$\kappa/T = a + bT^{\alpha-1}$ (refs. 35 and 36) with $\alpha$ fixed
to 2. The two terms $aT$ and $bT^{\alpha}$ represent electronic and
phonon contributions, respectively. In the phonon term, the value of
$\alpha$ is usually between 2 and 3, due to specular reflection of
phonons at the smooth crystal surfaces in the boundary scattering
limit at low temperature. \cite{Sutherland,SYLi} Previously,
$\alpha$ = 2.22 and 2.02 were observed in BaFe$_2$As$_2$
\cite{Kurita1} and BaFe$_{1.9}$Ni$_{0.1}$As$_2$ \cite{LDing} single
crystals, respectively. Here we only focus on the electronic term.

In zero field, the fitting gives residual linear term $\kappa_0/T$ =
-3 $\pm$ 9 $\mu$W K$^{-2}$ cm$^{-1}$. This value of $\kappa_0/T$ is
within the experimental error bar $\pm$ 5 $\mu$W K$^{-2}$ cm$^{-1}$,
\cite{SYLi} although the fitting error bar is a little high due to
the slight noise of the data. Even after considering these error
bars, the $\kappa_0/T$ is still less than 3\% of the normal-state
Wiedemann-Franz law expectation $L_0$/$\rho_0$ = 0.378 mW K$^{-2}$
cm$^{-1}$, with $L_0$ the Lorenz number 2.45 $\times$ 10$^{-8}$ W
$\Omega$ K$^{-2}$. Such a negligible $\kappa_0/T$ in zero field
suggests a nodeless (at least in $ab$-plane) superconducting gap in
overdoped BaFe$_{1.73}$Co$_{0.27}$As$_2$, which is consistent with
previous results on Ba$_{1-x}$K$_x$Fe$_2$As$_2$ \cite{XGLuo} and
BaFe$_{1.9}$Ni$_{0.1}$As$_2$, \cite{LDing} and different from
BaFe$_{1.86}$Co$_{0.14}$As$_2$. \cite{YMachida} It has been noted
that the large $\kappa_0/T$ in zero field observed in
BaFe$_{1.86}$Co$_{0.14}$As$_2$ by Machida {\it et al.}
\cite{YMachida} may be extrinsic. \cite{LTaillefer}

In $H$ = 9 and 14.5 T magnetic fields, $\kappa_0/T$ = 0.365 $\pm$
0.009 and 0.366 $\pm$ 0.009  mW K$^{-2}$ cm$^{-1}$ were obtaind from
the fittings, respectively. For both values, one gets the Lorenz
ratio $L = \rho_0\kappa_0/T$ = 0.97 $\pm$ 0.03$L_0$, which shows
that Wiedemann-Franz law is roughly satisfied within the
experimental error bar. Note that in the non-superconducting parent
BaFe$_2$As$_2$ single crystal, the Wiedemann-Franz law was found to
be satisfied as $T \rightarrow 0$. \cite{Kurita1}

The saturation of thermal conductivity in $H >$ 9 T suggests that
the bulk $H_{c2}$ has been reached at $H$ = 9 T, although the
resistive transition is not completely suppressed until $H_{c2}$ =
17 T. Similar situation happened in overdoped cuprate superconductor
Tl-2201 with $T_c$ = 15 K, in which $H_{c2}$ = 13 T was obtained
from the resistivity data, while bulk $H_{c2}$ = 7 T was determined
from the thermal conductivity data. \cite{Proust} Here we take bulk
$H_{c2}$ = 9 T for overdoped BaFe$_{1.73}$Co$_{0.27}$As$_2$. To
choose a slightly different bulk $H_{c2}$ does not affect our
discussions below.

In Fig. 3, the normalized $\kappa_0/T$ of
BaFe$_{1.73}$Co$_{0.27}$As$_2$ is plotted as a function of
$H/H_{c2}$, together with the clean $s$-wave superconductor Nb,
\cite{Lowell} the dirty $s$-wave superconducting alloy InBi,
\cite{Willis} the multi-band $s$-wave superconductor NbSe$_2$,
\cite{Boaknin} an overdoped sample of the $d$-wave superconductor
Tl-2201, \cite{Proust} and BaFe$_{1.9}$Ni$_{0.1}$As$_2$.
\cite{LDing} As seen in Fig. 3, the rapid increase of $\kappa_0/T$
at low field for overdoped BaFe$_{1.73}$Co$_{0.27}$As$_2$ is clearly
different from the optimally doped BaFe$_{1.9}$Ni$_{0.1}$As$_2$. In
fact, it looks more like the typical behavior of $d$-wave
superconductors, due to the Volovik effect. \cite{Volovik} However,
the negligible $\kappa_0/T$ in zero field, which means nodeless
superconducting gap, has excluded $d$-wave gap in
BaFe$_{1.73}$Co$_{0.27}$As$_2$.

\begin{figure}
\includegraphics[clip,width=7.2cm]{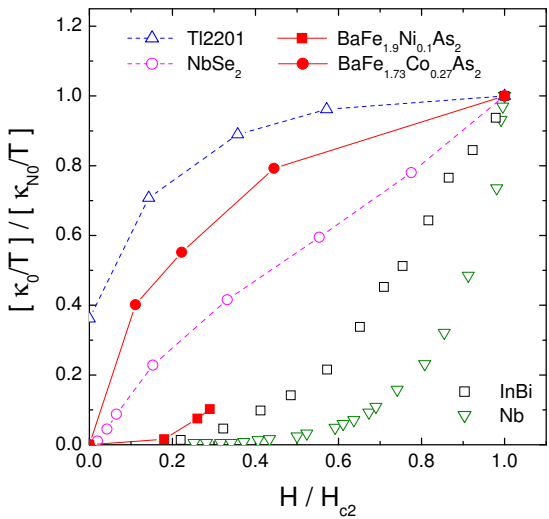}
\caption{(Color online) Normalized residual linear term $\kappa_0/T$
of BaFe$_{1.73}$Co$_{0.27}$As$_2$ as a function of $H/H_{c2}$.
Similar data of the clean $s$-wave superconductor Nb, \cite{Lowell}
the dirty $s$-wave superconducting alloy InBi, \cite{Willis} the
multi-band $s$-wave superconductor NbSe$_2$, \cite{Boaknin} an
overdoped sample of the $d$-wave superconductor Tl-2201,
\cite{Proust} and BaFe$_{1.9}$Ni$_{0.1}$As$_2$ \cite{LDing} are also
shown for comparison.}
\end{figure}

For $s$-wave superconductor NbSe$_2$, $\kappa_0/T$ is zero in $H$ =
0, but it also increases rapidly at low field, unlike Nb and InBi.
This has be explained by its multigap structure: the gap on the
$\Gamma$ band is approximately one third of the gap on the other two
Fermi surfaces, and magnetic field first suppresses the
superconductivity on the Fermi surface with smaller gap (given that
$H_{c2}(0) \propto \Delta_0^2$). \cite{Boaknin} Therefore, the even
sharper increase of $\kappa_0/T(H)$ in
BaFe$_{1.73}$Co$_{0.27}$As$_2$ may result from an extreme case of
multigap structure, in which the gap of one band is much smaller
than others (e.g., 1/4 or 1/5). However, there is an apparent
difference between NbSe$_2$ and BaFe$_{1.73}$Co$_{0.27}$As$_2$. For
NbSe$_2$, after the smaller gap was suppressed, $\kappa_0/T(H)$
shows a slight downward curvature at high field due to the larger
gap. Similar curvature of $\kappa_0/T(H)$ was found in MgB$_2$.
\cite{Sologubenko} In Fig. 3, $\kappa_0/T(H)$ of
BaFe$_{1.73}$Co$_{0.27}$As$_2$ increases so rapidly and it does not
show downward curvature at high field. Such an anomalous
$\kappa_0/T(H)$ has never been seen before. If it indeed results
from multiple nodeless gaps, the bands with smaller gaps must
contain most of the carriers so that $\kappa_0/T(H)$ can increase so
rapidly all the way to high field.

To investigate this possibility, we have examined the band structure
and superconducting gaps in doped BaFe$_2$As$_2$ system, revealed by
ARPES experiments. \cite{HDing,KNakayama,KTerashima,YSekiba} From
the hole doped side, in Ba$_{0.6}$K$_{0.4}$Fe$_2$As$_2$ ($T_c$ = 37
K), the average gap values $\Delta(0)$ for the two hole ($\alpha$
and $\beta$) pockets at the $\Gamma$ point are 12.5 and 5.5 meV,
respectively, while for the electron ($\gamma$ and $\delta$) pockets
at the $M$ point, the gap value is about 12.5 meV.
\cite{HDing,KNakayama} For electron-doped
BaFe$_{1.85}$Co$_{0.15}$As$_2$ ($T_c$ = 25.5 K) at optimal doping,
the inner hole ($\alpha$) pocket disappears, and the average gap
values $\Delta(0)$ of hole ($\beta$) and electron ($\gamma$ and
$\delta$) pockets are 6.6 and 5.0 meV, respectively.
\cite{KTerashima} Such nearly isotropic multigaps with similar size
has been used to explain the slow field dependence of $\kappa_0/T$
up to 30\% $H_{c_2}$ in BaFe$_{1.9}$Ni$_{0.1}$As$_2$, \cite{LDing}
as seen in Fig. 3. With further electron doping, in heavily doped
non-superconducting BaFe$_{1.7}$Co$_{0.30}$As$_2$, the $\beta$ hole
pocket is absent or very small, while the two electron ($\gamma$ and
$\delta$) pockets at the $M$ point significantly expand.
\cite{YSekiba} This is consistent with the band structure
calculation, which shows that the $\beta$ band disappears at
$\sim$26.5\% electron doping. \cite{DegangZhang} Based on this trend
of band structure evolution, in our superconducting
BaFe$_{1.73}$Co$_{0.27}$As$_2$ sample there should be a  very small
hole ($\beta$) pocket, together with two large electron ($\gamma$
and $\delta$) pockets which contain most of the carriers. To explain
its anomalous $\kappa_0/T(H)$, the gap on hole ($\beta$) pocket must
be much larger, 4 to 5 times, than the gaps on electron ($\gamma$
and $\delta$) pockets.

In the BCS theory, larger DOS usually leads to a larger
superconducting gap $\Delta(0)$. Therefore it is counterintuitive
that the vanishing $\beta$ pocket with much smaller DOS ends up with
a much larger gap. However, in the theory of interband
superconductivity, \cite{Dolgov} this is exactly the result of the
interband-only pairing, since the pairing amplitude on one band is
generated by the DOS on the other.

Right after the discovery of $T_c$ = 26 K superconductivity in
LaFeAsO$_{1-x}$F$_x$, \cite{Kamihara} the importance of interband
pairing interaction has been emphasized due to the multiband
fermiology. \cite{Mazin2,KKuroki} The interaction, possibly via
antiferromagnetic spin fluctuations, connects the well-separated
Fermi surface pockets located around $\Gamma$ and around $M$, and
gives extended $s$-wave pairing symmetry whose order parameter has
opposite sign on the electron and hole pockets.
\cite{Mazin2,KKuroki,FWang} Although the relative strength of
interband and intraband pairing interactions is still under debate,
\cite{Mazin2,KKuroki,FWang,KSeo,Dolgov} the experimentally observed
different $2\Delta/k_BT_c$ values on different Fermi surface pocket
in Ba$_{0.6}$K$_{0.4}$Fe$_2$As$_2$ and
BaFe$_{1.85}$Co$_{0.15}$As$_2$ prefer the interband pairing
mechanism. \cite{KTerashima}

Quantitatively, in the interband-only pairing model, the gap ratio
$\Delta_2/\Delta_1 = \sqrt{N_1/N_2}$, with $N_1$ and $N_2$ the
Fermi-level density of states. \cite{Dolgov} In this sense, our
overdoped BaFe$_{1.73}$Co$_{0.27}$As$_2$ sample has provdied the
best testing ground for the theory of interband superconductivity,
due to its biggest difference of DOS between the hole and electron
pockets in doped BaFe$_2$As$_2$ superconductors so far. Indeed, the
results of our current work have given strong support for the
interband superconductivity in FeAs-based superconductors. It will
be very interesting to directly measure the superconducting gaps in
our overdoped BaFe$_{1.73}$Co$_{0.27}$As$_2$ sample with ARPES,
which needs to be done at temperature below the $T_c$ = 8.1 K.

\section{Summary}

In summary, we have used low-temperature thermal conductivity to
clearly demonstrate nodeless superconducting gap in overdoped
iron-arsenide superconductor BaFe$_{1.73}$Co$_{0.27}$As$_2$.
Furthermore, the $\kappa_0/T(H)$ increases sharply at low field,
very different from Ba$_{1-x}$K$_x$Fe$_2$As$_2$ and
BaFe$_{1.9}$Ni$_{0.1}$As$_2$ near optimal doping. It may reveal an
exotic superconducting gap structure in overdoped
BaFe$_{1.73}$Co$_{0.27}$As$_2$: the vanishing hole ($\beta$) pocket
has a much larger gap than the electron ($\gamma$ and $\delta$)
pockets, although the electron pockets have much larger density of
states. Such an exotic gap structure is an evidence for the theory
of interband superconductivity, thus of great importance to
understand the superconducing state in FeAs-based superconductors.

\section{Note added}

During preparation of this manuscript (arXiv:0908.2209), a similar
work on BaFe$_{2-x}$Co$_x$As$_2$ was post online. \cite{Tanatar} In
ref. 52, the results of overdoped BaFe$_{1.772}$Co$_{0.228}$As$_2$
($T_c$ = 10.1 K) are consistent with ours, but the anomalous
increase of $\kappa_0/T(H)$ at low field was explained by highly
anisotropic superconducting gap with deep minima. While this debate
needs to be resolved by low-temperature ARPES experiments, we note a
very recent calculation of $\kappa_0/T(H)$ with unequal size of
isotropic $s_{\pm}$-wave gaps has successfully fit the experimental
data, \cite{YBang} thus supports our interpretation.

\begin{center}
{\bf ACKNOWLEDGEMENTS}
\end{center}

We thank Y. Chen, D. L. Feng, and J. P. Hu for useful discussions.
This work is supported by the Natural Science Foundation of China,
the Ministry of Science and Technology of China (National Basic
Research Program No:2009CB929203 and 2006CB601000), Program for New
Century Excellent Talents in University,
and STCSM of China (No: 08dj1400200 and 08PJ1402100).\\

$^*$ E-mail: shiyan$\_$li@fudan.edu.cn


\begin{thebibliography}{99}

\bibitem{Kamihara} Y. Kamihara, T. Watanabe, M. Hirano, and H. Hosono,  J. Am. Chem. Soc. {\bf 130}, 3296 (2008).
\bibitem{XHChen} X. H. Chen, T. Wu, G. Wu, R. H. Liu, H. Chen, and D. F. Fang, Nature {\bf 453}, 761 (2008).
\bibitem{GFChen} G. F. Chen, Z. Li, D. Wu, G. Li, W. Z. Hu, J. Dong, P. Zheng, J. L. Luo, and N. L. Wang, Phys. Rev. Lett. {\bf 100}, 247002 (2008).
\bibitem{ZARen} Z. A. Ren, W. Lu, J. Yang, W. Yi, X. L. Shen, C. Zheng, G. C. Che, X. L. Dong, L. L. Sun, F. Zhou, and Z. X. Zhao, Chin. Phys. Lett. {\bf 25}, 2215 (2008).
\bibitem{RHLiu} R. H. Liu, G. Wu, T. Wu, D. F. Fang, H. Chen, S. Y. Li, K. Liu, Y. L. Xie, X. F. Wang, R. L. Yang, L. Ding, C. He, D. L. Feng, and X. H. Chen, Phys. Rev. Lett. {\bf 101}, 087001 (2008).
\bibitem{KIshida} K. Ishida, Y. Nakai, and H. Hosono, J. Phys. Soc. Jpn. {\bf 78}, 062001 (2009).
\bibitem{Mazin1} I. I. Mazin and J. Schmalian, Physica C {\bf 469}, 614 (2009).
\bibitem{HDing} H. Ding, P. Richard, K. Nakayama, K. Sugawara, T. Arakane, Y. Sekiba, A. Takayama, S. Souma, T. Sato, T. Takahashi, Z. Wang, X. Dai, Z. Fang, G. F. Chen, J. L. Luo, and N. L. Wang,  EPL {\bf 83}, 47001 (2008).
\bibitem{TKondo} T. Kondo, A. F. Santander-Syro, O. Copie, C. Liu, M. E. Tillman, E. D. Mun, J. Schmalian, S. L. Bud'ko, M. A. Tanatar, P. C. Canfield, and A. Kaminski, Phys. Rev. Lett. {\bf 101}, 147003 (2008).
\bibitem{LZhao} L. Zhao, H. Y. Liu, W. T. Zhang, J. Q. Meng, X. W. Jia, G. D. Liu, X. L. Dong, G. F. Chen, J. L. Luo, N. L. Wang, W. Lu, G. L. Wang, Y. Zhou, Y. Zhu, X. Y. Wang, Z. Y. Xu, C. T. Chen, and X. J. Zhou, Chin. Phys. Lett. {\bf 25} 4402 (2008).
\bibitem{LWray} L. Wray, D. Qian, D. Hsieh, Y. Xia, L. Li, J. G. Checkelsky, A. Pasupathy, K. K. Gomes, C. V. Parker, A. V. Fedorov, G. F. Chen, J. L. Luo, A. Yazdani, N. P. Ong, N. L. Wang, and M. Z. Hasan, Phys. Rev. B {\bf 78} 184508 (2008).
\bibitem{KNakayama} K. Nakayama, T. Sato, P. Richard, Y. M. Xu, Y. Sekiba, S. Souma, G. F. Chen, J. L. Luo, N. L. Wang, H. Ding, and T. Takahashi, EPL {\bf 85}, 67002 (2009).
\bibitem{KTerashima} K. Terashima, Y. Sekiba, J. H. Bowen, K. Nakayama, T. Kawahara, T. Sato, P. Richard, Y.-M. Xu, L. J. Li, G. H. Cao, Z.-A. Xu, H. Ding, T. Takahashi, Proc. Natl. Acad. Sci. U.S.A. {\bf 106}, 7330 (2009).
\bibitem{TYChien} T. Y. Chen, Z. Tesanovic, R. H. Liu, X. H. Chen, and C. L. Chien, Nature {\bf 453}, 1224 (2008).
\bibitem{LShan} L. Shan, Y. L. Wang, X. Y. Zhu, G. Mu, L. Fang, C. Ren, and H. H. Wen, EPL {\bf 83} 57004 (2008).
\bibitem{Gonnelli} R. S. Gonnelli, D. Daghero, M. Tortello, G. A. Ummarino, V. A. Stepanov, J. S. Kim, and R. K. Kremer, Phys. Rev. B {\bf 79} 184526 (2009).
\bibitem{KMatano} K. Matano, Z. A. Ren, X. L. Dong, L. L. Sun, Z. X. Zhao, and G. Q. Zheng, EPL {\bf 83} 57001 (2008).
\bibitem{HGrafe} H. -J. Grafe, D. Paar, G. Lang, N. J. Curro, G. Behr, J. Werner, J. Hamann-Borrero, C. Hess, N. Leps, R. Klingeler, and B. B¨¹chner, Phys. Rev. Lett. {\bf 101}, 047003 (2008).
\bibitem{KAhilan} K. Ahilan, F. L. Ning, T. Imai, A. S. Sefat, R. Jin, M. A. McGuire, B. C. Sales, and D. Mandrus, Phys. Rev. B {\bf 78}, 100501(R) (2008).
\bibitem{MYashima} M. Yashima, H. Nishimura, H. Mukuda, Y. Kitaoka, K. Miyazawa, P. M. Shirage, K. Kiho, H. Kito, H. Eisaki, A. Iyo, J. Phys. Soc. Jpn. {\bf 78}, 103702
(2009).
\bibitem{CMartin1} C. Martin, M. E. Tillman, H. Kim, M. A. Tanatar, S. K. Kim, A. Kreyssig, R. T. Gordon, M. D. Vannette, S. Nandi, V. G. Kogan, S. L. Bud'ko, P. C. Canfield, A. I. Goldman, and R. Prozorov, Phys. Rev. Lett. {\bf 102}, 247002 (2009).
\bibitem{KHashimoto} K. Hashimoto, T. Shibauchi, T. Kato, K. Ikada, R. Okazaki, H. Shishido, M. Ishikado, H. Kito, A. Iyo, H. Eisaki, S. Shamoto, and Y. Matsuda, Phys. Rev. Lett. {\bf 102}, 017002 (2009).
\bibitem{LMalone} L. Malone, J. D. Fletcher, A. Serafin, A. Carrington, N. D. Zhigadlo, Z. Bukowski, S. Katrych, J. Karpinski, Phys. Rev. B {\bf 79}, 140501(R) (2009).
\bibitem{CMartin2} C. Martin, R. T. Gordon, M. A. Tanatar, H. Kim, N. Ni, S. L. Bud'ko, P. C. Canfield, H. Luo, H. H. Wen, Z. Wang, A. B. Vorontsov, V. G. Kogan, and R. Prozorov, Phys. Rev. B {\bf 80}, 020501(R) (2009).
\bibitem{RGordon} R. T. Gordon, N. Ni, C. Martin, M. A. Tanatar, M. D. Vannette, H. Kim, G. D. Samolyuk, J. Schmalian, S. Nandi, A. Kreyssig, A. I. Goldman, J. Q. Yan, S. L. Bud'ko, P. C. Canfield, and R. Prozorov, Phys. Rev. Lett. {\bf 102}, 127004 (2009).
\bibitem{TJWilliams} T. J. Williams, A. A. Aczel, E. Baggio-Saitovitch, S. L. Bud'ko, P. C. Canfield, J. P. Carlo, T. Goko, J. Munevar, N. Ni, Y. J. Uemura, W. Yu, G. M. Luke, arXiv:0905.3215.
\bibitem{Shakeripour} H. Shakeripour, C. Petrovic, and L. Taillefer, New J. Phys. {\bf 11}, 055065 (2009).
\bibitem{XGLuo} X. G. Luo, M. A. Tanatar, J.-Ph. Reid, H. Shakeripour, N. Doiron-Leyraud, N. Ni, S. L. Bud'ko, P. C. Canfield, H. Q. Luo, Z. S. Wang, H. H. Wen, R. Prozorov, L. Taillefer, Phys. Rev. B {\bf 80}, 140503(R) (2009).
\bibitem{LDing} L. Ding, J. K. Dong, S. Y. Zhou, T. Y. Guan, X. Qiu, C. Zhang, L. J. Li, X. Lin, G. H. Cao, Z. A. Xu, and S. Y. Li, New J. Phys. {\bf 11}, 093018 (2009).
\bibitem{YMachida} Y. Machida, K. Tomokuni, T. Isono, K. Izawa, Y. Nakajima, and T. Tamegai, J. Phys. Soc. Jpn. {\bf 78}, 073705 (2009).
\bibitem{JKDong} J. K. Dong, T. Y. Guan, S. Y. Zhou, X. Qiu, L. Ding, C. Zhang, U. Patel, Z. L. Xiao, and S. Y. Li, Phys. Rev. B {\bf 80}, 024518 (2009).
\bibitem{NKurita} N. Kurita, F. Ronning, Y. Tokiwa, E. D. Bauer, A. Subedi, D. J. Singh, J. D. Thompson, and R. Movshovich, Phys. Rev. Lett. {\bf 102}, 147004 (2009).
\bibitem{MYamashita} M. Yamashita, N. Nakata, Y. Senshu, S. Tonegawa, K. Ikada, K. Hashimoto, H. Sugawara, T. Shibauchi, and Y. Matsuda, Phys. Rev. B {\bf 80}, 220509(R) (2009).
\bibitem{LFang} L. Fang, H. Q. Luo, P. Cheng, Z. S. Wang, Y. Jia, G. Mu, B. Shen, I. I. Mazin, L. Shan, C. Ren, and H. H. Wen, Phys. Rev. B {\bf 80}, 140508 (R) (2009).
\bibitem{Sutherland} M. Sutherland, D. G. Hawthorn, R. W. Hill, F. Ronning, S. Wakimoto, H. Zhang, C. Proust, E. Boaknin, C. Lupien, and Louis Taillefer, Phys. Rev. B {\bf 67}, 174520 (2003).
\bibitem{SYLi} S. Y. Li, J.-B. Bonnemaison, A. Payeur, P. Fournier, C. H. Wang, X. H. Chen, and L. Taillefer, Phys. Rev. B {\bf 77}, 134501 (2008).
\bibitem{Kurita1} N. Kurita, F. Ronning, C. F. Miclea, E. D. Bauer, J. D. Thompson, A. S. Sefat, M. A. McGuire, B. C. Sales, D. Mandrus, and R. Movshovich, Phys. Rev. B {\bf 79}, 214439 (2009).
\bibitem{LTaillefer} Private communication with Dr. Louis Taillefer.
\bibitem{Proust} C. Proust, E. Boaknin, R. W. Hill, L. Taillefer, and A. P. Mackenzie, Phys. Rev. Lett. {\bf 89}, 147003 (2002).
\bibitem{Lowell} J. Lowell and J. Sousa, J. Low. Temp. Phys. {\bf 3}, 65 (1970).
\bibitem{Willis} J. Willis and D. Ginsberg, Phys. Rev. B {\bf 14}, 1916 (1976).
\bibitem{Boaknin} E. Boaknin, M. A. Tanatar, J. Paglione, D. Hawthorn, F. Ronning, R. W. Hill, M. Sutherland, L. Taillefer, Jeff Sonier, S. M. Hayden, and J. W. Brill, Phys. Rev. Lett. {\bf 90}, 117003 (2003).
\bibitem{Volovik} G. E. Volovik, JETP Lett. {\bf 58}, 469 (1993).
\bibitem{Sologubenko} A. V. Sologubenko, S. V. Borisenko, M. S. Golden, S. Legner, K. A. Nenkov, M. Knupfer, J. Fink, H. Berger, L. Forr¨®, and R. Follath, Phys. Rev. B {\bf 66}, 014504 (2002).
\bibitem{YSekiba} Y. Sekiba, T. Sato, K. Nakayama, K. Terashima, P. Richard, J. H. Bowen, H. Ding, Y-M. Xu, L. J. Li, G. H. Cao, Z-A. Xu, and T. Takahashi, New J. Phys. {\bf 11}, 025020 (2009).
\bibitem{DegangZhang} Degang Zhang, Phys. Rev. Lett. {\bf 103}, 186402 (2009).
\bibitem{Dolgov} O. V. Dolgov, I. I. Mazin, D. Parker, A. A. Golubov, Phys. Rev. B {\bf 79}, 060502(R) (2009).
\bibitem{Mazin2} I. I. Mazin, D. J. Singh, M. D. Johannes, and M. H. Du, Phys. Rev. Lett. {\bf 101}, 057003 (2008).
\bibitem{KKuroki} K. Kuroki, S. Onari, R. Arita, H. Usui, Y. Tanaka, H. Kontani, and H. Aoki, Phys. Rev. Lett. {\bf 101}, 087004 (2008).
\bibitem{FWang} F. Wang, H. Zhai, Y. Ran, A. Vishwanath, and D. H. Lee, Phys. Rev. Lett. {\bf 102}, 047005 (2009).
\bibitem{KSeo} K. Seo, B. Andrei Bernevig, and J. P. Hu, Phys. Rev. Lett. {\bf 101}, 206404 (2008).
\bibitem{Tanatar} M. A. Tanatar, J. P. Reid, H. Shakeripour, X. G. Luo, N. Doiron-Leyraud, N. Ni, S. L. Bud'ko, P. C. Canfield, R. Prozorov, Louis Taillefer, arXiv:0907.1276.
\bibitem{YBang} Yunkyu Bang, arXiv:0912.5049.

\end{thebibliography}
\end{document}